\documentclass[12pt]{article}

\usepackage[utf8]{inputenc}

\usepackage{amsmath, amssymb}
\usepackage{amsfonts}
\usepackage[framemethod=tikz]{mdframed}
\usepackage[french]{algorithm2e}
\usepackage{mathtools}
\usepackage[colorlinks]{hyperref}
\hypersetup{
  colorlinks=true,
  linkcolor=red,
  citecolor=red
}
\AtBeginDocument{%
  \hypersetup{%
    linkcolor=.,%
    citecolor=blue,%
  }%
}%

\makeatletter
\renewcommand{\@algocf@capt@plain}{above}
\makeatother

\begin{document}

\begin{center}
{\Large {\sc 
R\'egularisation dans les Mod\`eles Lin\'eaires G\'en\'eralis\'es Mixtes avec effet al\'eatoire autor\'egressif 
}}

\bigskip

Jocelyn Chauvet$^{1}$, Catherine Trottier$^{1,2}$ \& Xavier Bry$^{1}$
 
\bigskip

{\it
$^{1}$ Institut Montpelli\'erain Alexander Grothendieck, CNRS, Univ Montpellier, France.
jocelyn.chauvet@umontpellier.fr,  xavier.bry@univ-montp2.fr.

$^{2}$ Univ Paul-Val\'ery Montpellier 3, Montpellier, France. 
\\ 
catherine.trottier@univ-montp3.fr.
}
\end{center}

\bigskip

{\bf R\'esum\'e.} 
Nous proposons des versions régularisées de l'algorithme Espérance - Maxi-misation (EM) permettant d'estimer un Modèle Linéaire Généralisé Mixte (GLMM) pour des données de panel (mesurées sur plusieurs individus à différentes dates). 
Une réponse aléatoire $y$ est modélisée par un GLMM, au moyen d'un ensemble $X$ de variables explicatives et de deux effets aléatoires. 
Le premier effet modélise la dépendance des mesures relatives à un même individu, tandis que le second représente l'effet temporel autocorrélé partagé par tous les individus. 
Les variables dans $X$ sont supposées nombreuses et redondantes, si bien qu'il est nécessaire de régulariser la régression.
Dans ce contexte, nous proposons d'abord un algorithme EM pénalisé en norme $L_2$ pour des données de petite dimension, puis une version régularisée de l'algorithme EM, basée sur la construction de composantes supervisées, plutôt destinée à la grande dimension.

\smallskip

{\bf Mots-cl\'es.} Algorithme EM régularisé, Modèles Linéaires Généralisés Mixtes, Effet aléatoire autocorrélé, Données de panel.

\bigskip\bigskip

{\bf Abstract.} We address regularised versions of the Expectation-Maximisation (EM) algorithm for Generalised Linear Mixed Models (GLMM) in the context of panel data (measured on several individuals at different time points). A random response $y$ is modelled by a GLMM, using a set $X$ of explanatory variables and two random effects. The first effect introduces the dependence within individuals on which data is repeatedly collected while the second embodies the serially correlated time-specific effect shared by all the individuals. Variables in $X$ are assumed many and redundant, so that regression demands regularisation. 
In this context, we first propose a $L_2$-penalised EM algorithm for low-dimensional data, and then a supervised component-based regularised EM algorithm for the high-dimensional case.
\smallskip


\section{Introduction}
L'un des objectifs principaux de l'analyse des données de panel est de prendre en compte la dépendance engendrée par la présence de mesures répétées au cours du temps. 
Par ailleurs, au vu des facilités actuelles pour collecter de grandes masses de données, les fortes corrélations potentielles au sein des variables explicatives doivent également être considérées. Dans ce but, des régularisations de type ridge et lasso ainsi que des méthodes à composantes ont récemment été mises en avant.  

Dans le cadre des Modèles Linéaires Mixtes (LMM), 
\textcolor{blue}{\hyperlink{Eliot}{Eliot et al. (2011)}} 
proposent d'étendre la régression ridge aux données longitudinales. Afin de maximiser une vraisemblance pénalisée sur la norme $L_2$ des coefficients, ils suggèrent une variante de l'algorithme EM qui inclut, à chaque itération, la détermination du meilleur coefficient de pénalisation au travers d'une étape de Validation Croisée Généralisée (GCV).
Une autre méthode basée sur une vraisemblance pénalisée, cette fois-ci dans une perspective de sélection de variables, est proposé par 
\textcolor{blue}{\hyperlink{Schelldorfer}{Schelldorfer et al. (2014)}}. 
Ils élaborent à cet effet  un algorithme de type lasso pour ajuster des Modèles Linéaires Généralisés Mixtes (GLMM) de grande dimension, qui combine approximation de Laplace et algorithme de descente de gradient.

Dans le cadre des GLM, 
\textcolor{blue}{\hyperlink{Bry1}{Bry et al. (2013)}} 
mettent en œuvre une méthode de type PLS -- nommée Régression Linéaire Généralisée sur Composantes Supervisées (SCGLR)) -- qui régularise le prédicteur linéaire tout en facilitant son interprétation au moyen de composantes explicatives. 
Inspirés par l'algorithme de 
\textcolor{blue}{\hyperlink{Schall}{Schall (1991)}}, 
\textcolor{blue}{\hyperlink{Chauvet}{Chauvet et al. (2016)}} étendent cette stratégie de régularisation aux données groupées, et plus généralement à l'ensemble de la classe des GLMM.

À notre connaissance, les effets aléatoires apparaissant dans les stratégies précédentes sont supposés distribués selon des lois normales avec des niveaux indépendants. 
Cependant, pour les données de panel, il est naturel de greffer une structure d'autocorrélation à  l'effet aléatoire temporel.
Deux objectifs complémentaires émergent alors : étendre d'une part la régression ridge de 
\textcolor{blue}{\hyperlink{Eliot}{Eliot et al. (2011)}} 
aux GLMM avec effet aléatoire AR(1) ; et présenter d'autre part une nouvelle version de SCGLR adaptée pour les données de panel de grande dimension.  

\section{Modélisation}
Dans cette section, nous rappelons les hypothèses principales concernant les GLMM et nous introduisons les distributions des effets aléatoires.
Dans un souci de clarté, nous nous focaliserons sur des données de panel équilibrées avec $N$ individus, chacun d'eux observés en $T$ dates. On note $n = N \times T$ le nombre total d'observations, $X$ la matrice de design des effets fixes (de taille $n \times p$) et $U$ celle des effets aléatoires (de taille $n \times q$).
Par ailleurs, $Y$ désigne le vecteur de taille $n$ des réponses aléatoires, $\beta$ le vecteur de taille $p$ des effets fixes, et $\xi$ le vecteur de taille $q$ des effets aléatoires. Nous observons une réalisation $y$ de $Y$ tandis que $\xi$ n'est pas observé.
Nous supposons usuellement que :
\begin{itemize}
\item[(i)]
les $Y_i \, | \, \xi, \; i \in \left\lbrace 1, \ldots, n \right\rbrace$ 
sont indépendants et leurs distributions appartiennent à la famille exponentielle ; 
\item[(ii)]
l'espérance conditionnelle $\mu_i = \mathbb{E}(Y_i \, | \, \xi)$ 
dépend de  $\beta$ et $\xi$
au travers de la fonction de lien $g$ et du prédicteur linéaire 
$\eta_i = x_i^T\beta + u_i^T \xi$, vérifiant $\eta_i = g(\mu_i)$.
\end{itemize}
Dans notre modèle, nous considérons deux effets aléatoires $\xi_1$ et $\xi_2$, aux rôles et distributions bien différents :
\begin{itemize}
\item[(i)]
$\xi_1$ est l'effet aléatoire spécifique aux individus. En les supposant indépendants, on pose :
$$
\xi_1 \sim \mathcal{N}_N \left( 0, \, \sigma_1^2 I_N \right),
$$
où $\sigma_1^2$ est la composante ``individuelle'' de la variance, supposée inconnue.
\item[(ii)]
$\xi_2$ est l'effet aléatoire temporel partagé par l'ensemble des individus, ce dernier pouvant être vu comme un phénomène latent non pris en considération dans les variables explicatives. Ayant tendance à perdurer au cours du temps, on le modélise à l'aide d'un processus autorégressif d'ordre 1 (AR(1)), i.e. pour tout $t \in \left\lbrace 1, \ldots, T-1 \right\rbrace$,
\begin{align*}
\xi_{2,t+1} &= \rho \, \xi_{2,t} + \nu_t, \\ 
\nu_t &\overset{\text{iid}}{\sim} 
\mathcal{N}( 0, \, \sigma_2^2),
\end{align*}
où $\rho$ est le paramètre de l'AR(1) et $\sigma_2^2$ la composante ``temporelle'' de la variance, supposés inconnus.
De tels effets temporels latents apparaissent naturellement par exemple dans un contexte économique (où les agents partagent la même politique et conjoncture économiques dont les effets ont une certaine inertie temporelle), 
ou bien en biologie (car l'environnement écologique est souvent trop complexe pour être observé de manière exhaustive au travers des variables explicatives).    
\end{itemize}
Enfin, $\xi_1$ and $\xi_2$ sont supposés indépendants. 
En notant
${\xi=\left( \xi_1^T, \xi_2^T \right)^T}$, ${U_1=I_N \otimes \textbf{1}_T}$,
${U_2=\textbf{1}_N \otimes I_T}$ and ${U=\left[ U_1 \, | U_2 \right]}$, 
le prédicteur linéaire $\eta$ peut être écrit matriciellement : 
$$ \eta = X\beta + U\xi.$$

\section{Méthodes}
En raison de la structure de dépendance des GLMM, l'algorithme des scores de Fisher a été adapté par 
\textcolor{blue}{\hyperlink{Schall}{Schall (1991)}} 
afin d'estimer le modèle.
Dans le but de tenir compte à la fois des fortes redondances dans $X$ ainsi que des distributions non-conventionnelles des effets aléatoires, nous envisageons dans la suite la possibilité d'introduire une étape de type EM régularisé au sein d'un algorithme de Schall.
Chaque itération se décompose alors en deux étapes clés : une étape de linéarisation et une étape d'estimation régularisée.

\medskip

{\bf \underline{Étape de linéarisation}.}
Pour tout $i \in \left\lbrace 1, \ldots, n \right\rbrace$, la linéarisation à l'ordre 1 de $y_i$ au voisinage de $\mu_i$ est donnée par :
$g(y_i) \simeq z_i = g(\mu_i) + (y_i - \mu_i) g'(\mu_i)$.
Matriciellement, cette approximation fournit une variable dite ``de travail'' $z$ s'exprimant au travers du modèle linéarisé suivant
$$  \mathcal{M} : \quad
z = X\beta + U\xi + e,
$$
avec $\text{Var}(e \,|\, \xi) = \text{Diag} 
\left( 
\left[ g'(\mu_i) \right]^2 \text{Var}(Y_i \,|\, \xi)
\right)_{i=1, \ldots, n} = \Gamma$.

\medskip

{\bf \underline{Étape d'estimation}.}
Au lieu de résoudre le système de Henderson associé à $\mathcal{M}$ vu comme un LMM (à la manière de 
\textcolor{blue}{\hyperlink{Schall}{Schall (1991)}} ), 
nous proposons plutôt une étape de type EM régularisé. 
Pour des données de petite dimension ($p<n$), nous suggérons d'étendre l'algorithme EM avec pénalité ridge élaboré par 
\textcolor{blue}{\hyperlink{Eliot}{Eliot et al. (2011)}}. 
Par contre, dans le cas $p \gg n$, nous lui préférons un algorithme EM régularisé basé sur la construction de composantes supervisées.

\subsection{Données de petite dimension}
Notre étape d'estimation prend appui sur 
\textcolor{blue}{\hyperlink{Green}{Green (1990)}}, 
qui popularise l'utilisation de l'algorithme EM lorsque la vraisemblance est pénalisée, et 
\textcolor{blue}{\hyperlink{Golub}{Golub et al. (1979)}}, 
qui encouragent l'utilisation de la GCV pour choisir efficacement le coefficient de pénalisation 
$\lambda$.
Cependant, contrairement au LMM homoscédastique considéré par 
\textcolor{blue}{\hyperlink{Eliot}{Eliot et al. (2011)}}, 
$\mathcal{M}$ contient des erreurs hétéroscédastiques.
Nous optons alors plutôt pour le critère GCV proposé par 
\textcolor{blue}{\hyperlink{Andrews}{Andrews (1991), p. 372}},  
cohérent avec les modèles hétéroscédastiques. 
En notant $\theta = \left( \beta, \sigma_1^2, \sigma_2^2, \rho \right)$, 
nous présentons l'itération générique de notre algorithme EM pénalisé adapté aux GLMM avec effet aléatoire AR(1) dans l'
\textcolor{red}{\autoref{ridge_GLMM_AR}} ci-dessous.

\subsection{Données de grande dimension}
Dans le cas $p \gg n$, au lieu de maximiser une fonction objectif $\mathcal{Q}_{\text{pen}}$ pénalisée par la norme $L_2$ des coefficients, nous explorons la possibilité de maximiser une fonction 
$\mathcal{Q}_{\text{reg}}$ régularisée à l'aide de composantes. Pour une unique composante $f$, elle s'écrit sous la forme :
\begin{align*}
\mathcal{Q}_{\text{reg}} \left(\theta, \theta^{[t]} \right) &= 
\mathbb{E}_{\xi|z}
\left[
L_{\text{reg}}(\theta ; z,\xi) \,|\, \theta^{[t]}
\right], \, \text{avec} \\
L_{\text{reg}}(\theta ; z,\xi) &= (1-s) L(\theta ; z,\xi) + s \phi(w),
\end{align*}
où $\phi(w)$ est un critère de pertinence structurelle (PS) introduit par \textcolor{blue}{\hyperlink{Bry2}{Bry et Verron (2015)}} et 
$s \in \left[ 0,1 \right]$ un paramètre permettant de régler l'importance relative de la PS par rapport à $L$, vue ici comme une mesure de la qualité d'ajustemeent.
Avec $l \geqslant 1$, $\phi(w)$ s'écrit : 
$$
\phi(w) = \left(
\sum_{j=1}^p \left[ \text{cor}^2 \left( x^j, f \right) \right]^l
\right)^{\frac{1}{l}}.
$$
Pour des raisons d'identifiabilité, la composante s'écrit $f=Cw$, avec $C=XU$ 
l'ensemble des composantes principales de $X$ de valeurs propres non-nulles.
Les paramètres $s$ et $l$ sont calibrés par validation croisée et les composantes de rangs supérieurs sont calculées comme celle de rang 1, 
après l'ajout de contraintes d'orthogonalité aux précédentes.
\bigskip

\vfill
\begin{algorithm}[H]
\DontPrintSemicolon
\caption{Itération générique de l'algorithme EM pénalisé en norme $L_2$ pour GLMM avec effet aléatoire AR(1).}
\label{ridge_GLMM_AR}
\begin{mdframed}[
    linecolor=black,
    linewidth=2pt,
    roundcorner=4pt,
    backgroundcolor=olive!15,
    userdefinedwidth=\textwidth,
]
\begin{itemize}
\item[\bf (1)] {\bf \underline{Étape de linéarisation}.}
Définir le modèle linéarisé par :
$$
\mathcal{M}^{[t]} : z^{[t]} = X\beta + U\xi + e, \; \text{avec} \,  
\text{Var}(e \,|\, \xi) = \Gamma^{[t]}.  
$$
\item[\bf (2)] {\bf \underline{Étape d'estimation}.}
\begin{itemize}
\item[\bf (2.a)]
$L$ désignant la log-vraisemblance complétée du modèle linéarisé, \\
définir la log-vraisemblance complétée pénalisée $L_{\text{pen}}$ par :  
$$L_{\text{pen}}(\theta ; z,\xi) \vcentcolon = L(\theta ; z,\xi) - \frac{\lambda}{2} \beta^T \beta$$
\item[\bf (2.b)]
Avec $\widehat{z}^{[t]}$ les valeurs ajustées et 
$S_{\lambda}^{[t]}$ la ``hat-matrix'' 
vérifiant l'égalité \\ $\widehat{z}^{[t]} = S_{\lambda}^{[t]} z^{[t]}$, 
poser : 
$$
\lambda^{[t]} \longleftarrow \text{arg} \; \underset{\lambda}{\min} \left\lbrace
\text{GCV}(\lambda) = 
\dfrac{n^{-1} \left\lVert z^{[t]} - S_{\lambda}^{[t]} z^{[t]} 
\right\rVert^2_{{\Gamma^{[t]}}^{-1}}}
{\left[  1 - n^{-1} \text{tr} \left( S_{\lambda}^{[t]} \right)  \right]^2}
\right\rbrace.
$$
\item[\bf (2.c)] \textbf{Étape E.} Calculer : 
$$
\mathcal{Q}_{\text{pen}} \left(\theta, \theta^{[t]} \right) \vcentcolon = \mathbb{E}_{\xi|z}
\left[
L_{\text{pen}}(\theta ; z^{[t]}, \xi) \,|\, \theta^{[t]}, \lambda^{[t]}
\right].
$$
\item[\bf (2.d)] \textbf{Étape M.} Poser alors : 
$$
\theta^{[t+1]} \longleftarrow \text{arg } \underset{\theta}{\max} \;
\mathcal{Q}_{\text{pen}} \left(\theta, \theta^{[t]} \right).
$$
\end{itemize}
\item[\bf (3)] {\bf \underline{Mise à jour}.}
Poser $\xi^{[t+1]} = \mathbb{E}_{\xi|z} \left( \xi \,|\, \theta^{[t+1]} \right)$, et mettre à jour la variable de travail $z^{[t+1]}$ ainsi que la matrice de variance-covariance $\Gamma^{[t+1]}$.
\end{itemize}
\bigskip
Les étapes {\bf (1)--(3)} 
sont répétées tant que la stabilité conjointe des paramètres $\beta$, 
$\sigma_1^2$, $\sigma_2^2$ et $\rho$ n'est pas observée.
\end{mdframed}
\end{algorithm}
\vfill

\newpage
\section{Résultats numériques}
Afin d'évaluer les performances des deux méthodes, nous présenterons des études sur données simulées, notamment dans le cas Poisson - lien log. Ces simulations auront trois objectifs principaux :
\begin{itemize}
\item[(i)]
juger du nombre d'itérations nécessaire à la stabilisation des paramètres estimés, et ainsi se faire une idée de la vitesse de convergence des algorithmes proposés, 
\item[(ii)]
s'assurer que les MSE relatifs à chacun des paramètres convergent bien vers 0 lorsque la taille du jeu de données augmente, 
\item[(iii)]
vérifier que les méthodes proposées se comportent de manière identique quelle que soit la valeur de $\rho \in \left] -1,1 \right[$.
\end{itemize}

\section*{Bibliographie}

\hypertarget{Andrews}{
\noindent [1] Andrews, D.W. (1991). 
        Asymptotic optimality of generalized CL, cross-validation, and
        generalized cross-validation in regression with heteroskedastic
        errors. 
        {\it Journal of Econometrics}, 
        {\bf 47}, 359\,--\,377.}
        
\medskip    

\hypertarget{Bry1}{    
\noindent [2] Bry, X., Trottier, C., Verron, T. et Mortier, F. (2013).
       Supervised component generalized linear regression using a pls-extension
       of the fisher scoring   
       algorithm. 
       {\it Journal of Multivariate Analysis}, 
       {\bf 119}, 47\,--\,60. } 
        
\medskip  

\hypertarget{Bry2}{  
\noindent [3] Bry, X. et Verron, T. (2015). 
THEME: THEmatic model exploration through multiple co-structure maximization. 
{\it Journal of Chemometrics}, {\bf 29}, 637\,--\,647. }   

\medskip

\hypertarget{Chauvet}{           
\noindent [4] Chauvet, J., Trottier, C., Bry, X. et Mortier, F. (2016). $\,$
       Extension $\,$ to $\,$ mixed models of the Supervised Component-based
       Generalised Linear Regression. 
       {\it COMPSTAT: Proceedings in Computational Statistics}.}
        
\medskip    

\hypertarget{Eliot}{         
\noindent [5] Eliot, M., Ferguson, J., Reilly, M.P. et Foulkes, A.S. (2011).
       Ridge Regression for Longitudinal Biomarker Data.
       {\it The International Journal of Biostatistics}, 
       {\bf 7}, 1, Article 37.}
        
\medskip    

\hypertarget{Golub}{         
\noindent [6] Golub, G.H., Heath, M. et Wahba, G. (1979). 
       Generalized cross - validation  as a method for choosing a good ridge parameter. 
       {\it Technometrics}, 
       {\bf21}, 215\,--\,223.}
        
\medskip    

\hypertarget{Green}{         
\noindent [7] Green, P.J. (1990). 
       On use of the EM for penalized likelihood estimation. 
       {\it Journal of the Royal Statistical Society, Series B}, 
       {\bf 52}, 443\,--\,452.}
        
\medskip
    
\hypertarget{Schall}{         
\noindent [8] Schall, R. (1991). 
       Estimation in generalized linear models with random effects. 
       {\it Biometrika}, 
       {\bf 78}, 719\,--\,727.}
        
\medskip    

\hypertarget{Schelldorfer}{         
\noindent [9] Schelldorfer, J., Meier, L. et B\"{u}hlmann, P. (2014). 
       Glmmlasso: an algorithm for high-dimensional generalized linear mixed models using $l_1$-penalization.    
       {\it Journal of Computational and Graphical Statistics}, 
       {\bf 23}, 460\,--\,477.}
\end{document}